\documentclass[reqno,foot]{amsart}
\usepackage{graphicx}
\usepackage{epstopdf}
\usepackage{subcaption}
\usepackage[flushleft]{threeparttable}
\usepackage{multirow}
\usepackage{hyperref}
\usepackage[sort&compress,square,numbers]{natbib}
\usepackage{amsaddr}

\title[Precision spectroscopy of helium in a magic wavelength optical...]{Precision spectroscopy of helium in a magic wavelength optical dipole trap}
\author[R.J.~Rengelink \textit{et al.}]{R.J.~Rengelink\textsuperscript{1}}
\author{Y.~van~der ~Werf\textsuperscript{1}}
\author{R.P.M.J.W.~Notermans\textsuperscript{1,\textdagger}}
\author{R.~Jannin\textsuperscript{1}}
\author{K.S.E.~Eikema\textsuperscript{1}}
\author{M.D.~Hoogerland\textsuperscript{2}}
\author{W.~Vassen\textsuperscript{1,\textasteriskcentered}}
\address[1]{LaserLaB, Department of Physics and Astronomy, Vrije Universiteit, De Boelelaan 1081, 1081 HV Amsterdam, the Netherlands}
\address[2]{Dodd-Walls Centre for Photonic and Quantum Technologies, Department of Physics, University of Auckland, Private Bag 92019, Auckland, New Zealand}
\address[\textdagger]{Present address: Department of Physics, Stanford University, Stanford, California 94305, USA}
\address[\textasteriskcentered]{Corresponding author: \MakeLowercase{w.vassen@vu.nl}}

\begin{document}
\begin{abstract}
Improvements in both theory and frequency metrology of few-electron systems such as hydrogen and helium have enabled increasingly sensitive tests of quantum electrodynamics (QED), as well as ever more accurate determinations of fundamental constants and the size of the nucleus. At the same time advances in cooling and trapping of neutral atoms have revolutionized the development of increasingly accurate atomic clocks. Here, we combine these fields to reach the highest precision on an optical tranistion in the helium atom to date by employing a Bose-Einstein condensate confined in a magic wavelength optical dipole trap. The measured transition accurately connects the ortho- and parastates of helium and constitutes a stringent test of QED theory. In addition we test polarizability calculations and ultracold scattering properties of the helium atom. Finally, our measurement probes the size of the nucleus at a level exceeding the projected accuracy of muonic helium measurements currently being performed in the context of the proton radius puzzle.
\end{abstract}

\maketitle

\clearpage

In the past decades, high-precision spectroscopy measurements in atomic physics scale systems have pushed precision tests of quantum electrodynamics (QED), one of the cornerstones of the standard model of physics, ever further~\cite{Parthey11,Pachucki17} and have led to accurate determinations of fundamental constants~\cite{Mohr16,Hanneke08,Biraben09,Sturm14}. Recently however, measurements of transition frequencies in muonic hydrogen ($\mathrm{\mu H}$) have revealed a discrepancy of six standard deviations~\cite{Pohl10,Antognini13} with respect to the accepted CODATA value for the proton charge radius. This discrepancy, which has become known as the ``proton radius puzzle", has stimulated strong interest in the field, as its confirmation implies the violation of lepton universality, one of the pillars of the standard model. New experiments in atomic hydrogen~\cite{Beyer17,Fleurbaey18}, and muonic deuterium~\cite{Pohl16} have only deepened the puzzle, prompting research into other elements such as muonic helium ($\mathrm{\mu \ ^{3,4}He^+}$)~\cite{Nebel12}. From these measurements the charge radii of the alpha-particle and the helion (1.68~fm resp. 1.97~fm) are projected to be determined with sub-attometer accuracy, which should be compared to high-precision experiments in electronic helium atoms or ions.

QED theory of the helium atom, with two electrons more complicated than hydrogen, has seen impressive improvements in recent years, with QED corrections up to order $m \alpha^6$ now evaluated~\cite{Pachucki17}. Recent experiments are in good agreement~\cite{Kandula10,Hodgman09,CancioPastor04,Luo13,Noter14a,Luo16,Zheng17b,Huang18} and may allow a competitive value for the fine structure constant in the near future~\cite{Borbely09,Smiciklas10,Zheng17a,Marsman15}. The anticipated evaluation of the next highest order corrections ($m \alpha^7$)~\cite{Pachucki17} would allow the determination of the $^4$He nuclear charge radius with an accuracy better than 1\%. At present nuclear charge radii can already be determined differentially, i.e. with respect to $\mathrm{^4 He}$, due to cancellation of higher-order terms in the isotope shift. Using this approach the radii of the exotic halo nuclei $\mathrm{^6He}$ and $\mathrm{^8He}$~\cite{Wang04,Mueller07}, as well as the stable isotope $\mathrm{^3 He}$~\cite{Shiner95,CancioPastor12,vRooij11} were determined with accuracies far exceeding electron scattering experiments~\cite{Sick14}. However, different experiments on the $^3$He-$^4$He isotope shift show significant discrepancies~\cite{Pachucki17}, even between different measurements of the same dipole allowed $2 \ ^3 S_1 \rightarrow 2 \ ^3 P$ transition~\cite{Zheng17b}. Furthermore, improving the experimental accuracy on this transition is challenging due to the 1.6~MHz natural linewidth and the presence of quantum interference shifts~\cite{Marsman15}. Only one previous experiment has used the doubly forbidden $2 \ ^3 S_1 \rightarrow 2 \ ^1 S_0$ transition~\cite{vRooij11}, which in contrast has an excellent quality factor of $2.4 \times 10^{13}$ (natural linewidth 8~Hz) that poses no fundamental limit in the foreseeable future.

Here we report a new measurement of the $2 \ ^3 S_1 \rightarrow 2 \ ^1 S_0$ transition frequency at 1557~nm which improves the previous result by an order of magnitude, making this the most accurate optical frequency measurement in the helium atom to date ($\delta \nu / \nu = 1.0 \times 10^{-12}$). Our measurement has been performed using a Bose-Einstein condensate (BEC) in the metastable $2 \ ^3 S_1$ state confined in an optical dipole trap (ODT) at a previously predicted~\cite{Noter14b} magic wavelength for this transition. At such a magic wavelength the ac-Stark shift on the transition vanishes, a property that has been exploited to realize atomic clocks operating at a stability in the $10^{-19}$ region~\cite{Campbell17,Marti18}, allowing constraints on a possible time-variation of fundamental constants~\cite{Ludlow15}. Moreover, ab-initio calculations of polarizability have recently emerged as an alternative means of testing atomic theory at a level sensitive to QED effects~\cite{Mitroy13,Zhang16,Henson15}. By finetuning the ODT laser wavelength to reduce the ac-Stark shift on the transition frequency, we measure the magic wavelength to high accuracy, providing a stringent test for ab-initio calculations~\cite{Wu18}.

Our approach has also enabled us to measure the mean-field, or cold-collision, shift on the transition for the first time by direct observation. This frequency shift was instrumental in the first observation of Bose-Einstein condensation of atomic hydrogen via two-photon excitation of the $1 S \rightarrow 2 S$ transition, where the enormous density of the BEC gave rise to a huge mean-field shift~\cite{Fried98, Killian98}. The associated transition lineshape allowed quantitative analysis of these results~\cite{Killian00}. In earlier work~\cite{Noter16}, we showed how this lineshape is affected by an asymmetry in the trapping potential for $2 \ ^3 S_1$ and $2 \ ^1 S_0$ atoms, and we could extract the $2 \ ^1 S_0 - 2 \ ^3 S_1$ scattering length with 50\% accuracy. Now, working in a magic wavelength trap, we are able to improve this accuracy by an order of magnitude.

These measurements therefore test our knowledge of the helium atom in three different ways. The transition frequency measured here is a test of level energies and is sensitive to the finite size of the nucleus. The magic wavelength determination is a precision test of atomic structure as a whole and is therefore also sensitive to transition dipole moments. Finally, the scattering length derived from the mean-field shift is a precise test of the molecular potentials between helium atoms.

\section*{Setup}
\begin{figure}[h]
\centering
\hspace{-1cm}
\includegraphics[width= \textwidth]{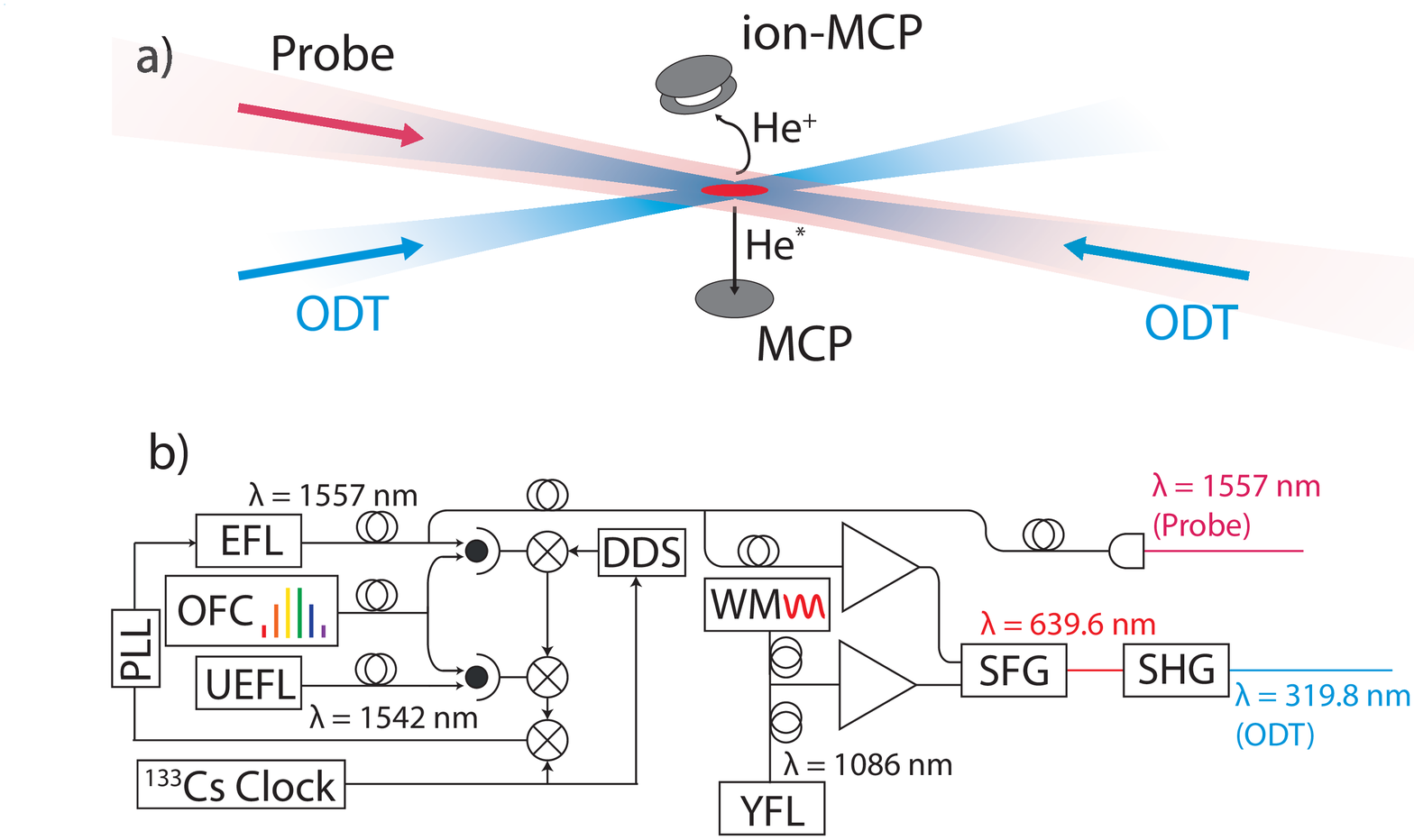}
\caption{a) Schematic of the experimental geometry. Two overlapping laser beams crossing at an angle of $19^\circ$ form the ODT. The probe light is counterpropagating with one of the ODT beams. A high-voltage biased MCP detector above the setup detects ions generated by excited atoms. A grounded MCP below the setup detects the remaining metastable atoms that fall under gravity when they are released from the trap. b) Schematic of the laser setup. An erbium fiber laser (EFL) is transfer-locked in a phase locked loop (PLL) to an ultrastable erbium fiber laser (UEFL) via an optical frequency comb (OFC). Control over the frequency offset is provided by an in-loop direct digital synthesizer (DDS). The EFL serves as the probe laser, but part of it is also split off to seed a fiber amplifier. An independent ytterbium fiber laser (YFL) is amplified and overlapped with this light in order to generate the sum frequency (SFG), which is frequency doubled in a second harmonic generation (SHG) stage. A wavemeter (WM) is used to measure the wavelength of the YFL.}
\label{fig:setup}
\end{figure}

\begin{figure}[h]
\centering
\includegraphics[width= 0.75 \textwidth]{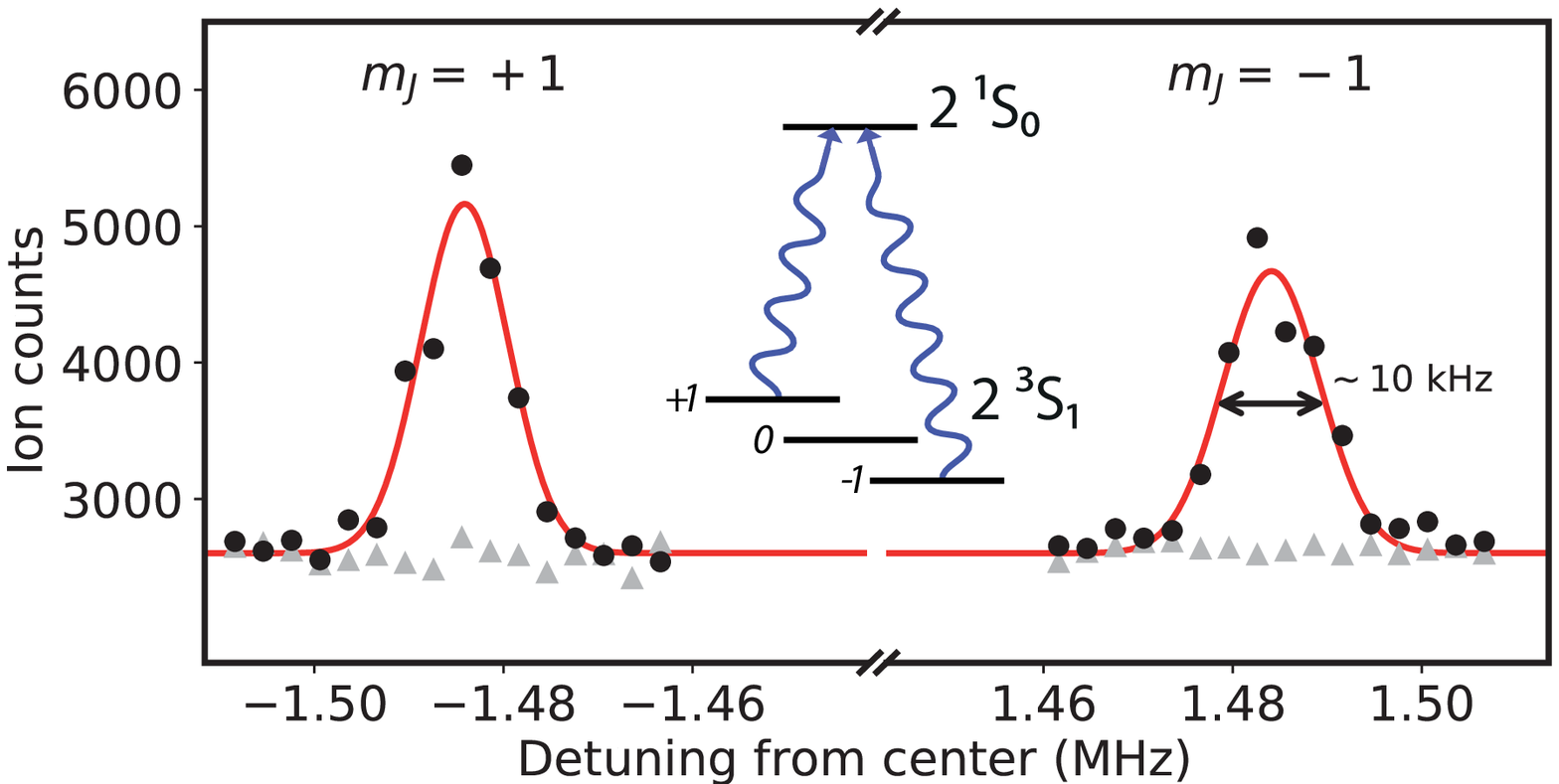}
\caption{A typical spectroscopy scan. The black circles indicate the signal when spectroscopy light is applied. The grey triangles are background calibration points measured directly after each spectroscopy point. In order to account for the Zeeman shift, the atoms are alternately excited from the $m_J = +1$, and $m_J=-1$ states. The red line is a fit of two Gaussians showing typical widths of about 10~kHz.}
\label{fig:typscan}
\end{figure}
We prepare a BEC of typically $10^6$ atoms in the metastable $2 \ ^3 S_1$ state (19.82 eV above the $1 \ ^1 S_0$ ground state, lifetime $\sim8000 \ \mathrm{s}$~\cite{Hodgman09})~\cite{vRooij11}, and transfer it into a dipole trap at 319.8~nm. The atoms are spin-polarized in the spin-stretched $m_J=+1$ state so that ionization via two-body collisions (Penning ionization) is strongly suppressed~\cite{Vassen12}. Figure~\ref{fig:setup}a shows the geometry of the dipole trap. A tightly focused ODT beam is passed through the vacuum chamber, refocused and passed through the chamber again with orthogonal linear polarization, intersecting itself at an angle of $19^\circ$. The atoms are trapped at the intersection, where the probe laser is applied counterpropagating to the incident ODT beam. To detect excitation of the transition, we measure the increased Penning ionization rate from the excited $2 \ ^1S_0$ atoms using a microchannel plate detector (MCP) and counter (see Methods). This detection method provides substantially better signal-to-noise ratio compared to a signal based on the loss of $2 \ ^3 S_1$ atoms used previously~\cite{vRooij11,Noter14a,Noter16}. After excitation, the remaining atoms (\textgreater 90\%) are dropped under gravity on another MCP detector placed 17~cm directly below the trap, producing a time-of-flight (TOF) signal. From a bimodal fit to the TOF signal, we determine the chemical potential and atom number of the BEC, as well as the temperature and atom number of the thermal cloud.

Fig.~\ref{fig:setup}b shows the optical setup generating the probe and trap laser light. Part of the probe laser light is also amplified and mixed with a second independent laser in order to generate the ODT light. This second laser is monitored by a high resolution wavemeter to determine the trap laser wavelength (see Methods). The optical and electronic setup for generating the probe and ODT laser light are described in refs.~\cite{Noter16} and~\cite{Rengelink16}.

To account for the Zeeman shift arising from the ambient magnetic field in the laboratory, we alternate between exciting from the $m_J=+1$ and $m_J=-1$ state (see Methods), which have first-order Zeeman shifts of equal magnitude but opposite sign. Exciting from the $m_J=0$ state, which shows no first-order Zeeman shift, is not possible due to a high Penning ionization rate~\cite{Vassen12}. Every measurement is alternated with a background measurement in order to monitor the level of background ion counts. In this way, we build up a double-peak spectrum as shown in fig.~\ref{fig:typscan}. We fit each measured line with two Gaussian peaks (see Methods) and calculate the center frequencies.

\section*{Results}

\begin{figure}[h]
\centering
\includegraphics[width= \textwidth]{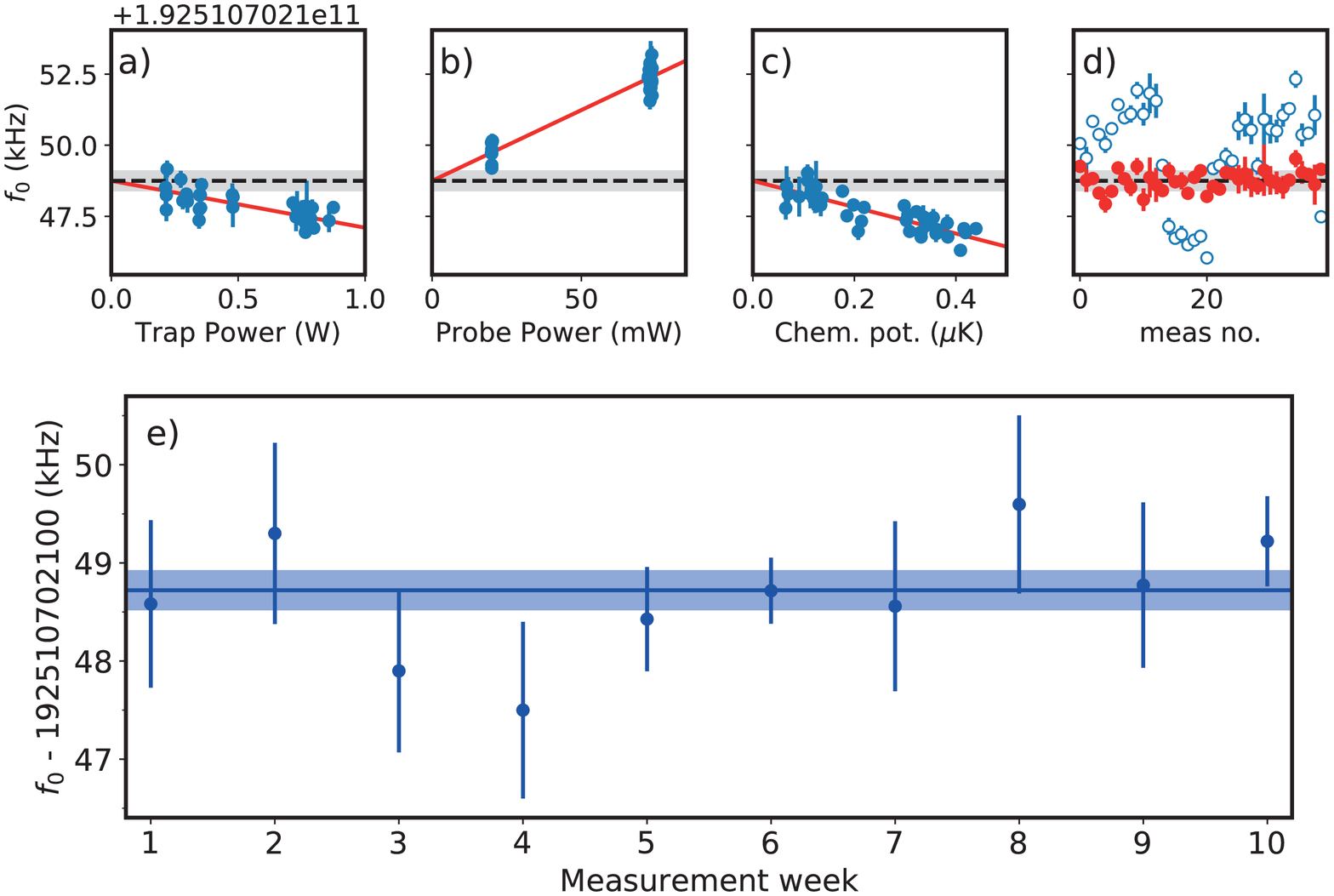}
\caption{a-d) Results of a multiple regression fit to a single dataset. The graphs a) through c) are the partial residual plots for each of the fit parameters. The final graph d) shows the measured frequencies (blue open circles) and the residuals of the regression model (red circles). The grey band in all these figures indicates the 1$\sigma$ uncertainty on the transition frequency determined from this particular dataset. e) Measured transition frequency per measurement week including systematic errors. The data point at week six is derived from the multiple regression fit shown in a-d, and the other points are based on similar datasets. The blue line and blue band indicate the weighted average and 1$\sigma$ uncertainty.}
\label{fig:allsum}
\end{figure}

By employing a magic wavelength ODT, the ac-Stark shift induced by the trap is greatly reduced compared to previous work~\cite{vRooij11}. The magic wavelength was not known with sufficient accuracy to eliminate the ac-Stark shift completely, and a residual trap-induced ac-Stark shift remains as a systematic shift that needs to be calibrated. In addition to this, two other systematic shifts are present that contribute roughly equally to the final accuracy: the ac-Stark shift from the probe laser, and the mean-field shift which is proportional to the chemical potential of the BEC.

In order to account for these systematics we performed multiple measurements in which we varied the ODT and probe laser powers as well as the chemical potential of the BEC. Since all of these systematic shifts are linear with respect to their corresponding experimental observable, we can fit the data with a multiple linear regression model, as shown in fig.~\ref{fig:allsum}. From this model we extracted the transition frequency as well as the slopes of the ac-Stark shifts and the mean-field shift simultaneously. For every measurement week, a single complete fit of the regression model was performed, where the total number of measured transition frequencies varied between 16 and 39.

It was experimentally not possible to vary all parameters independently. In particular the trap power and chemical potential are highly correlated because a deeper trap is better able to hold a high number of atoms at high density. To break this correlation as much as possible, we varied the chemical potential of the BEC while keeping the trap power fixed. This was achieved by varying the hold time in the ODT before applying probe light between 200 ms and a few seconds. Due to the fairly short ($\sim 2$ seconds, limited by off-resonant scattering of the ODT light) one-body lifetime of the BEC in the ODT, this allows for significant modification of the size of the BEC.

\subsection*{Magic wavelength}
\begin{figure}[h]
\centering
\includegraphics[width= \textwidth]{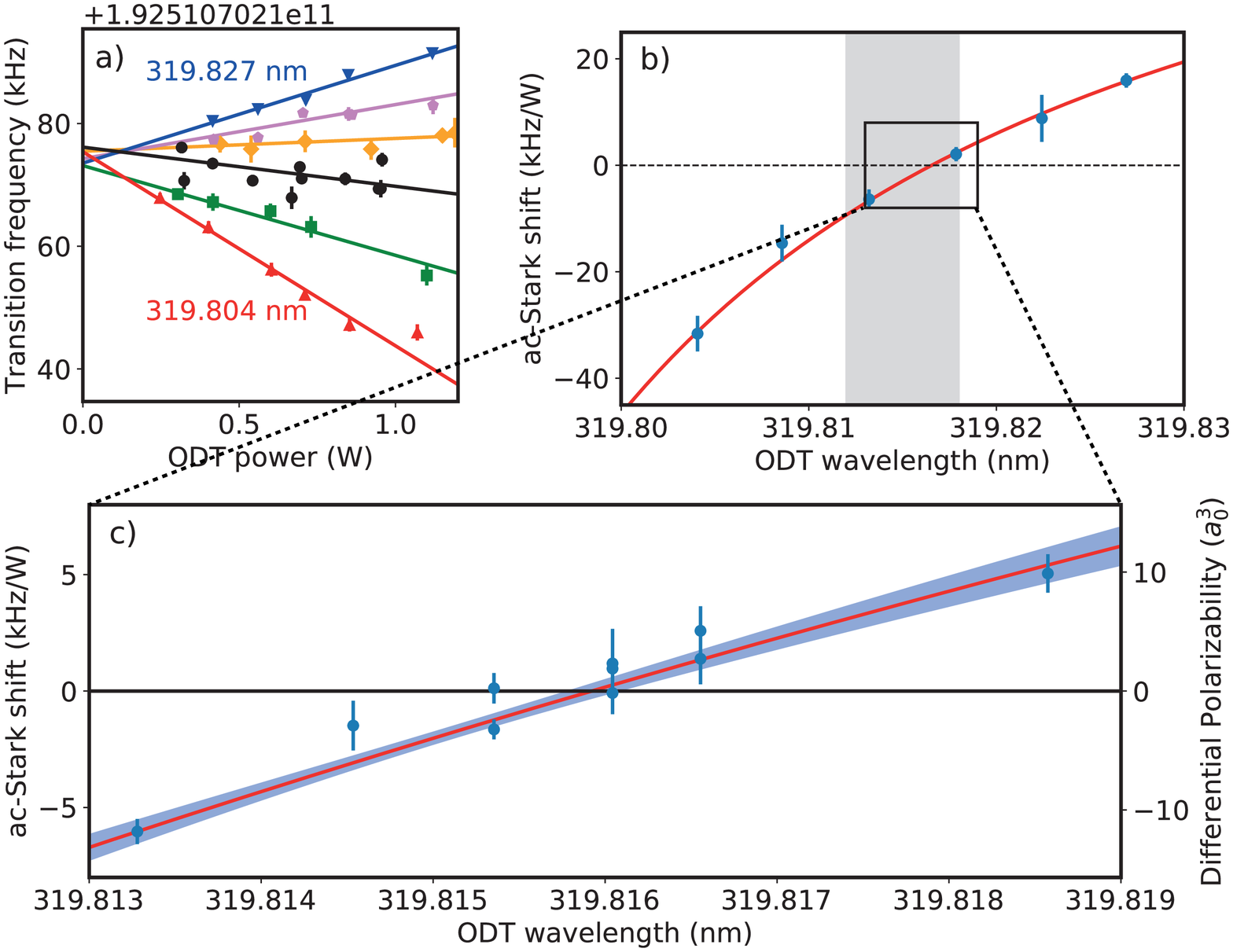}
\caption{a) Transition frequency as a function of laser power, showing only the ``Coarse" scan, with a linear fit at each wavelength in order to determine the ac-Stark shift. b) Slopes of the fitted lines in a) as a function of ODT laser wavelength. The gray region indicates the predicted range for the magic wavelength~\cite{Noter14b}. c) ``Fine" scan of the ac-Stark shift with the polarizability curve from ref.~\cite{Noter14b} fitted (red line). The blue region represents the 1$\sigma$ uncertainty on the fit. The magic wavelength condition is found at the zero-crossing.}
\label{fig:mwvl}
\end{figure}

The determination of the magic wavelength was performed over two measurement campaigns: a first ``coarse" campaign, and a second ``high resolution" campaign during which the absolute transition frequency was also measured. The results from the coarse campaign are shown in figs.~\ref{fig:mwvl}a,b, along with the predicted range from calculations~\cite{Noter14b}. These measurements gave a first estimate of the magic wavelength. However, during this first campaign the mean-field shift was not corrected for, leading to a small systematic offset on the magic wavelength.

During the final measurement campaign, the trap laser wavelength was varied over a smaller range in order to more precisely pinpoint the magic wavelength. Fig.~\ref{fig:mwvl}c shows the results from this campaign. In order to determine the magic wavelength we compare these data to the calculated polarizability curves~\cite{Noter14b}. The dominant uncertainty in these calculations depends only very weakly on wavelength and appears as a constant offset. The calculated polarizability can be related directly to the measured Stark shift by a scaling factor which corrects for the laser intensity (see Methods). By fitting the data to the calculated polarizability with a constant offset and a scaling factor, we also account for the slight curvature of the polarizability curve over the measured wavelength range. From this fit we then extract the laser intensity, a constant offset correction to the calculated polarizability, and the zero-crossing. We find the light intensity at the center of the trap to be $1.0(1) \times 10^8 \ \mathrm{W m^{-2}}$ using a 1~W ODT beam. This intensity is roughly half of our estimate assuming perfect focusing conditions and beam quality~\cite{Rengelink16}. The constant offset correction to the polarizability is found to be 3.4(5) $a_0^3$. The zero crossing of the fitted curve corresponds to the magic wavelength and is found at 319.815 92(15)~nm, which is in good agreement with the calculated value of 319.815(3)~nm. The vector and tensor part of the polarizability are negligible at the current level of uncertainty (see supplementary material), and also do not influence the measurement of the transition frequency.

\subsection*{Mean-field shift and scattering length}
The excited $2 \ ^1 S_0$ atoms experience a different mean-field potential compared to the remaining $2 \ ^3 S_1$ atoms because of the difference in scattering length. This leads to a shift of the transition frequency known as the mean-field, or cold-collision shift~\cite{Killian98}. The full-width of the mean-field lineshape $S(\nu)$~\cite{Killian00} turns out to be small compared to the observed linewidth. At the maximum density used in the experiments (peak density $n(0) \approx 4.5 \times 10^{13} \mathrm{cm^{-3}}$ or equivalently $\mu \approx k_B \times 0.5 \ \mathrm{\mu K}$), a full width $\delta \nu_{max} \approx 4.4 \ \mathrm{kHz}$ is expected (see supplementary material). The additional ac-Stark shift contribution to the width~\cite{Noter16} is negligible for the range of ODT laser wavelengths used in the final measurement campaign. Possible line-pulling effects due to the asymmetric lineshape  were investigated by fitting Gaussians to simulations of the broadened lineshape but were not found to affect the fitted frequencies.

The only observable effect of the mean-field interaction is therefore the average shift of this lineshape. We derive this average shift analytically by integrating the shift over the lineshape (see supplementary material for a detailed derivation):
\begin{equation}
\langle \Delta \nu_{MFS} \rangle = \frac{\int \nu S(\nu) d\nu}{\int S(\nu) d\nu} = \frac{4}{7 h} \left( \frac{a_{tt} - a_{ts}}{a_{tt}} \right) \mu,
\label{eq:MFS}
\end{equation}
where $a_{tt}$ and $a_{ts}$ are the scattering lengths for triplet-triplet and triplet-singlet collisions respectively, and $\mu$ is the chemical potential of the BEC.

The mean-field shift slope was found by including a linear regression to the chemical potential of the BEC in the multiple regression model shown in fig.~\ref{fig:allsum}c. Averaging over all measurements, we find a slope of -5.0(4) $\mathrm{kHz \ \mu K^{-1}}$. By rewriting equation~\ref{eq:MFS} we can express the unknown triplet-singlet scattering length in units of the very well known triplet-triplet scattering length, $a_{tt} = +7.512(5) \ \mathrm{nm} = +142.0(1) \ a_0$~\cite{Moal06}.  We find $a_{ts} = +82.5(5.2) \ a_0$, which is in agreement with our previous result of $a_{ts} = +50(10)_{stat}(43)_{syst} \ a_0$~\cite{Noter16}.

\subsection*{Transition frequency}
The final measured transition frequency is corrected for a number of systematic shifts as shown in table~\ref{tab:Err_and_corr}. By far the largest of these is the recoil shift correction due to the absorption of a 1557~nm photon, $\Delta f_{rec} = -h/(2 m \lambda^2) = -20.554 \ \mathrm{kHz}$, with negligible uncertainty.

\begin{table}

\begin{threeparttable}
\caption{Measured $2 \ ^3 S_1 \rightarrow 2 \ ^1 S_0$ transition frequency along with corrections. The final result is compared to several alternative determinations. Values are in kHz.}
\begin{tabular}[h]{| l r r |}
\hline
\bf{Term} & \bf{Correction} & \bf{Uncertainty} \\
\hline
Measured frequency & 192 510 702 169.31\hspace{5pt} & \\
Recoil shift & -20.554 & \\
ac-Stark shift: Probe & & \multirow{3}{*}{$\Bigg\rbrace$ 0.192*}\\
ac-Stark shift: ODT & & \\
Mean-field shift & & \\
Statistical & & 0.032\hspace{5pt} \\
Cs clock & -0.036 & 0.010\hspace{5pt} \\
Black-body radiation shift & \textless 0.005 & \\
dc-Stark shift & \textless 0.001 & \\
Quantum interference & \textless $10^{-4}$ & \\
Second-order Zeeman & \textless $10^{-5}$ & \\
\hline
Total: & 192 510 702 148.72\hspace{5pt} & 0.20\hspace{10pt} \\ 
van Rooij \textit{et al.}~\cite{vRooij11} & 192 510 702 145.6\hspace{10pt} & 1.8\hspace{15pt} \\
IE($2 \ ^3 S_1$) - IE($2 \ ^1 S_0$)~\cite{Huang18,Luo16,Zheng17b,Morton06} & 192 510 702 156\hspace{18pt} & 42\hspace{23pt} \\
Pachucki \textit{et al.}~\cite{Pachucki17} (theory) & 192 510 703 400\hspace{18pt} & 800\hspace{23pt} \\
\hline
\end{tabular}
\begin{tablenotes}
\small
\item * Uncertainty is correlated in the multiple regression model.
\end{tablenotes}
\label{tab:Err_and_corr}
\end{threeparttable}
\end{table}

Another systematic effect affecting all measurements equally is the frequency offset of the Cesium clock with respect to the SI-second to which all measurements are referenced. By comparing the clock with GPS time over the course of the entire measurement campaign, a fractional frequency offset of $-1.9(2) \times 10^{-13}$ was found (see Methods). By correcting for this offset, the clock was calibrated to within its specified stability floor of $5 \times 10^{-14}$, which contributes to the error budget. The measured transition frequency was corrected for the clock offset, corresponding to -36~Hz on the optical frequency.

Additional systematics are the black-body radiation shift, a dc-Stark shift due to the ion-MCP bias voltage, possible shifts due to quantum interference with far-off resonant transitions~\cite{Marsman15}, and the second-order Zeeman shift. None of these contribute significantly to the final error budget, and could be neglected in the final result. Details of these estimations can be found in the supplementary material.

Figure~\ref{fig:allsum}e shows the weekly average of all frequency measurements corrected for the systematic effects identified. Averaging over all results, we find a $2 \ ^3 S_1 \rightarrow 2 \ ^1 S_0$ transition frequency of 192 510 702 148.72(20) kHz, which corresponds to a relative uncertainty $\delta \nu / \nu = 1.0 \times 10^{-12}$.

\section*{Discussion and Conclusion}
The magic wavelength found in this work is in very good agreement with our earlier calculation~\cite{Noter14b} but is more accurate by over an order of magnitude. Very recent full-configuration-interaction calculations incorporating relativistic and recoil effects give the magic wavelength as 319.816~07(9)~nm, which is of a similar accuracy as our measurement and in excellent agreement~\cite{Wu18}. It is interesting to make the comparison with measurements on the tune-out wavelength (the wavelength for which the polarizability vanishes) for the $2 \ ^3 S_1$ level at 413~nm~\cite{Henson15}. Here a discrepancy with high precision calculations was found which was attributed to QED effects~\cite{Zhang16}, indicating that measurements of atomic polarizablity can be used as an alternative means of testing QED. 

The triplet-singlet scattering length $a_{ts}$ derived from the mean-field shift measured in this work is more accurate than the previous experimental bound~\cite{Noter16} by an order of magnitude, and in good agreement. This value can be used to test quantum chemistry calculations of the relevant molecular potentials. Interestingly, a previously reported estimate, derived from ab-initio calculations of the $1 \ ^3 \Sigma^+_g$ and $2 \ ^3 \Sigma^+_g$ molecular potentials~\cite{Mueller91}, found $a_{ts} = +42^{+0.5}_{-2.5} \ a_0$~\cite{Noter16}, which disagrees significantly with the value of $a_{st} = +82.5(5.2) \ a_0$ found in this work. This discrepancy may be related to the high ionization cross section which causes the complex optical potential method used in these calculations to break down.

The $2 \ ^3 S_1 \rightarrow 2 \ ^1 S_0$ transition frequency  obtained in this work improves the earlier measurement by van Rooij \textit{et al.}~\cite{vRooij11} by an order of magnitude. The results differ by 1.7$\sigma$ (see table~\ref{tab:Err_and_corr}). This difference may be due to a slight underestimation of the mean-field shift in that work, which was reported as negligible at the level of 1.1~kHz. Based on the slope of the mean-field shift found in this work and a rough estimation of the chemical potentials used in ref.~\cite{vRooij11}, we estimate that the mean-field shift in that work may have been somewhat larger (up to 2~kHz), which brings the results to within 1$\sigma$ of each other.

We can test for consistency with other experiments by taking the difference in ionization energy (IE) between the $2 \ ^3 S_1$ and $2 \ ^1 S_0$ levels. These IEs can be determined from transition frequency measurements~\cite{Huang18,Luo16, Zheng17b} and the theoretical IE of the $3 \ ^{1,3}D$ levels~\cite{Morton06} which can be calculated to high accuracy. As shown in table~\ref{tab:Err_and_corr}, the measured transition frequency is in excellent agreement with this difference though more accurate by more than two orders of magnitude. The measured transition frequency is also in reasonable agreement (1.6$\sigma$) with direct QED calculation~\cite{Pachucki17}, although the estimated uncertainty in this calculation is several orders of magnitude larger (see table~\ref{tab:Err_and_corr}). This uncertainty is currently of the same order as the total nuclear size shift, but is anticipated to be reduced~\cite{Pachucki17} which would allow a direct determination of the $^4$He nuclear charge radius from the measured transition frequency.

At the current state of the theory, nuclear size information can still be derived at high accuracy by looking at the $^3$He-$^4$He isotope shift on this transition, for which the estimated uncertainty in the calculations is much smaller (0.19~kHz)~\cite{Pachucki17}. Taking the difference between the transition frequency measured in this work and the $2 \ ^3 S_{1,F=3/2} \rightarrow 2 \ ^1 S_{0,F=1/2}$ transition frequency in $\mathrm{^3He}$~\cite{vRooij11}, we derive an updated value of the differential nuclear charge radius $\delta r^2 = r^2(\mathrm{^3 He}) - r^2(\mathrm{^4 He}) = 1.041(7) \ \mathrm{fm^2}$ (see ref.~\cite{Pachucki17} for details of the calculation), where the error is now dominated by the 1.5~kHz accuracy on the $\mathrm{^3 He}$ transition frequency. This new value agrees with electron scattering ($\delta r^2 = 1.066\pm0.06 \ \mathrm{fm^2}$~\cite{Sick14}) but still disagrees with determinations based on the $2 \ ^3 S_1 \rightarrow 2 \ ^3 P_{0,1,2}$ transitions~\cite{Shiner95,CancioPastor12}. The very recent measurement of the $2 \ ^3 S_{1} \rightarrow 2 \ ^3 P_{1}$ transition frequency in $\mathrm{^4He}$~\cite{Zheng17b} showed a 20$\sigma$ discrepancy with ref.~\cite{CancioPastor12}, which also indicates the need for further investigation of that transition. In the immediate future, we aim to improve the measurement of the $2 \ ^3 S_{1,F=3/2} \rightarrow 2 \ ^1 S_{0,F=1/2}$ transition in $\mathrm{^3He}$, which may bring the uncertainty on $\delta r^2$ down to $<0.002 \ \mathrm{fm^2}$. This is actually better than the expected accuracy from muonic helium, which will be limited to 0.0031~fm$^2$ because of theoretical uncertainty in calculating the two-photon exchange contribution to the Lamb shift~\cite{Diepold16}.  

The measurements presented in this work push our knowledge of the helium atom at several levels. The $2 \ ^3 S_1 \rightarrow 2 \ ^1 S_0$ transition frequency is measured to be 192~510~702~148.72(20)~kHz, intimately tying the ortho- and parastates together and allowing us to extract the $\mathrm{^3He} - \mathrm{^4He}$ nuclear charge radius difference with improved accuracy. The magic wavelength on this transition is determined to be 319.815 92(15)~nm, in good agreement with calculations and provides a stringent test for precision calculations of polarizabilities. Finally, the measurement of the mean-field shift allows extraction of the $2 \ ^3 S_1 - 2 \ ^1 S_0$ scattering length as +82.5(5.2) $a_0$, which disagrees significantly with recent quantum-chemistry calculations.

\section*{Acknowledgements}
We would like to thank Ruud van der Beek for useful discussions and a critical reading of the manuscript, Daniel Cocks and Ian Whittingham for helpful discussions, and Rob Kortekaas for technical support. We gratefully acknowledge financial support from the Netherlands Organisation for Scientific Research (NWO).

\section*{Author contributions}
R.J.R. and R.P.M.J.W.N. constructed the experimental setup, R.J.R., Y.W. and M.D.H. performed the measurements, R.J.R., Y.W., and R.J. investigated systematic effects and R.J.R. performed the data analysis. R.J.R., R.P.M.J.W.N. and K.S.E.E. performed and discussed the frequency metrology. W.V. initiated and supervised the project. All authors discussed the results and contributed to the manuscript.

\section*{Competing interests}
The authors declare no competing interests.

\section*{Data availability}
Experimental data is available from the corresponding author upon reasonable request.

\section*{Methods}
\subsection*{Experimental sequence}
The tightly focused (waist$<100 \ \mathrm{\mu m}$) ODT beams trap the atoms in a cigar-shaped harmonic potential. At 1~W input power, typical trap frequencies are $\omega_{ax} \approx 2 \pi \times 35 \ \mathrm{Hz}$ in the axial direction, and $\omega_{rad} \approx 2 \pi \times 300 \ \mathrm{Hz}$ in the radial direction. The probe laser beam has an input power of up to $80 \ \mathrm{mW}$ and a larger beam waist ($\sim 300 \ \mu m$) in order to ensure uniform illumination. We align the probe beam by overlapping it with the incident ODT beam. The polarization of the probe beam is linear but the direction is rotated with a motorized rotation stage in order to optimize the ion signal depending on whether the transition is made from the $m_J=+1$ or  the $m_J=-1$ state.

In the ODT, probe light is applied for about 100~ms, after which the remaining atoms are released to fall under gravity onto the MCP detector. This excitation time is chosen to yield sufficient signal while being short enough to not alter the chemical potential of the BEC by more than a few percent. During this step, the excited $2 \ ^1 S_0$ atoms collide with the remaining $2 \ ^3 S_1$ atoms in a strongly Penning ionizing collision channel. We expect an ionization rate comparable to that of unpolarized $2 \ ^3 S_1$ atoms, corresponding to a lifetime of about 1~ms for the $2 \ ^1 S_0$ atoms. The $\mathrm{He^+}$ ions produced by this process are detected by a second MCP detector (the ion-MCP) biased at -2.5 kV and located 11~cm above the trap. The signal from the ion-MCP is amplified by a pulse amplifier/discriminator and passed into a counter to yield the spectroscopy signal. Based on an excitation fraction of $\sim 5\%$, and a peak signal height of a few thousand counts from a BEC of a few million atoms, we estimate a detection efficiency of $\sim 2 \%$. We attribute the low efficiency to shielding of the trap volume by the grounded re-entrant windows.

To mitigate the Zeeman shift, we alternate between exciting from the $m_J=+1$ and $m_J=-1$ state. We transfer the atoms from the $m_J=+1$ to the $m_J=-1$ state via a Landau-Zener sweep which consists of a magnetic field ramp while RF-coupling between the magnetic substates is applied~\cite{Borbely12}. After application of the probe light a second sweep brings the $m_J=-1$ atoms back to the $m=+1$ state so that the TOF is not affected. In the case of $m_J=+1$ atoms the same sweep is performed without the RF-coupling to make sure no systematic magnetic field difference is introduced. After every measurement an identical measurement is performed with the probe light blocked in order to calibrate the level of background ion counts. We attribute this background to ionization of background gas by the 320~nm ODT light. This is corroborated by the fact that the background count rate is linearly proportional to the ODT laser power and increases when the background pressure is increased by temporarily closing the safety valve going to the main chambers' turbopump.

We estimate the linewidth of the probe laser at about 5~kHz, based on the combined effects of residual frequency comb noise, electronic noise on the phase-locked loop, and the 60~meter uncompensated fiber link between the frequency comb and the setup. The observed lineshapes are broader however, showing approximately Gaussian profiles of about 10~kHz width. We attribute the additional broadening to a small residual oscillation of the BEC inside the trap which causes Doppler broadening. Absorption images of the expanding BEC indeed show random velocity fluctuations with a standard deviation of about 3-4~mm/s in the axial direction. A simple model of a damped harmonic oscillator driven by statistical fluctuations of the axial trap position can quantitatively explain these observations.

\subsection*{Polarizability and ac-Stark shift}
The main uncertainty in the calculations of ref.~\cite{Noter14b} is due to approximations made in estimating the contribution to the polarizability due to coupling to the ionization continuum. Since this contribution is far off-resonant, we can neglect its wavelength dependence and treat it as a dc-offset. The calculated polarizabilities are given in atomic units which can be converted into SI using $\left[ 1 \ a_0^3 \right]_{a.u.} = \left[ 4 \pi \epsilon_0 a_0^3 \right]_{SI} \approx 1.64877 \times 10^{-41} \mathrm{J V^{-2} m^2}$. The intensity of the laser beam can now be calculated from the scale of the polarizability compared to the ac-Stark shift using $I = 2 \epsilon_0 c h \Delta \nu / \mathrm{Re} (\Delta \alpha)$, where $\Delta \alpha$ is the differential polarizability, and $\Delta \nu$ is the observed ac-Stark shift~\cite{Grimm00}.

\subsection*{Absolute frequency determination}
The ODT laser wavelength is derived from both the spectroscopy laser and a second free-running fiber laser. Because the spectroscopy laser frequency is determined with much higher accuracy, the uncertainty on the ODT laser wavelength is dominated by the free-running laser. This wavelength is measured continuously during the course of the experiment using a high resolution wavemeter (High Finesse WSU-30) with a specified accuracy of 30~MHz, which was periodically calibrated on the $2 \ ^3 S_1 \rightarrow 2 \ ^3 P_2$ line at 1083~nm. The laser wavelength was manually adjusted using the temperature control whenever it drifted by more than 50~MHz from the wavelength setpoint for that measurement week.  

The spectroscopy laser is locked to an ultrastable laser at 1542~nm (Menlo systems) via an optical frequency comb to bridge the wavelength gap in a transfer-lock configuration~\cite{Noter16}. The ultrastable laser serves as stable short term flywheel oscillator for the measurement. Over the course of a measurement day, the frequency of this reference is measured with respect to the Cs clock.

In order to reconstruct the absolute frequencies of the lasers several beatnotes are continuously  measured with a zero-dead time frequency counter, referenced to the Cs clock. The frequencies which are measured are the frequency comb carrier offset frequency, the down-mixed pulse repetition rate, the spectroscopy laser beat-note (before mixing in the DDS), and either the virtual beat-note or the ultrastable laser beat note. The wavelengths of the lasers were measured using a wavemeter with sufficient resolution to determine the comb modenumber of the observed beat notes.

From these data the ultrastable laser frequency was reconstructed, and a linear fit allows us to compensate for the slow drift of this laser during the day. This drift was found to be 22(2) mHz/s ($1.1(1) \times 10^{-16} \  \mathrm{s^{-1}}$) on average and fluctuating from day to day with a standard deviation of 9 mHz/s. The modified Allan deviation of these data agrees well with the specified stability of the Cs clock at the measured time scales, typically reaching a stability in the low $10^{-13}$ region after a single measurement day. In total, the spectroscopy data were acquired over about 30 separate measurement days, yielding in total about $5 \times 10^5$ seconds of total integration time, which is enough to reach the clock's stability floor of $5\times10^{-14}$.

During the full measurement campaign the time delay between the Cs clock and GPS pulse per second signal was continuously measured. The Allan deviation of this delay averages down as $\tau^{-1}$, and catches up with the Cs clock stability after about $10^6 \ \mathrm{s}$. Integrating over the full course of the measurement campaign, which took several months ($\sim 8 \times 10^6 \ \mathrm{s}$), we observed a fractional frequency drift of $-1.9(2) \times 10^{-13}$, with an accuracy that exceeds the specified Cs clock stability. We corrected for this drift in the frequency measurement data but take the specified clock stability floor as a conservative estimate of the uncertainty. The deviations between the GPS disseminated second with respect to the SI definition as reported in the BIPM circular T bulletin~\cite{BIPMcircT} were found to be negligible at the current level of uncertainty.
\subsection*{Data processing and statistical analysis}
The measured transition frequency data are fit with a weighted linear least squares model regressing to the trap and probe laser powers (measured before and after each scan) and the chemical potential of the BEC (as determined from the MCP time-of-flight profile). In order to separate the purely statistical error from the error due to the systematic shifts, we calculate the point of minimum uncertainty from the covariance matrix of the fit. At this point the uncertainty on the transition frequency is not correlated to the uncertainty in the other parameters and can be considered purely statistical, amounting to 32~Hz. Extrapolating from this point to zero laser power and chemical potential is associated with a systematic uncertainty of 192~Hz, which constitutes the bulk of the uncertainty in this work.

\bibliographystyle{naturemag}

\begin{thebibliography}{99}
\bibitem{Parthey11}
Parthey, C. G., \textit{et al.}, Improved Measurement of the Hydrogen 1S\textendash 2S Transition Frequency. \textit{Phys. Rev. Lett.} \textbf{107}, 203001 (2011)

\bibitem{Pachucki17}
Pachucki, K., Patk\'o\v s, V. and Yerokhin, V. A. Testing fundamental interactions on the helium atom. \textit{Phys. Rev. A} \textbf{92}, 062510 (2017)

\bibitem{Mohr16}
Mohr, P. J. Taylor, B. N. and Newell, D.B. CODATA recommended values of the fundamental physical constants: 2014. \textit{Rev. Mod. Phys.} \textbf{88}, 035009 (2016)

\bibitem{Hanneke08}
Hanneke, D., Fogwell, S. and Gabrielse, G. New Measurement of the Electron Magnetic Moment and the Fine Structure Constant. \textit{Phys. Rev. Lett.}  \textbf{100} 120801 (2008)

\bibitem{Biraben09}
Biraben, F. Spectroscopy of atomic hydrogen, How is the Rydberg constant determined? \textit{Eur. Phys. J. Special Topics} \textbf{172}, 109-119 (2009)

\bibitem{Sturm14}
Sturm, S. \textit{et al.} High-precision measurement of the atomic mass of
the electron. \textit{Nature} \textbf{506}, 467-470 (2014)

\bibitem{Pohl10}
Pohl, R. \textit{et al.} The size of the proton. \textit{Nature} \textbf{466}, 213-216 (2010)

\bibitem{Antognini13}
Antognini, A. \textit{et al.} Proton Structure from the
Measurement of 2S-2P Transition
Frequencies of Muonic Hydrogen. \textit{Science} \textbf{339}, 417-420 (2013)

\bibitem{Beyer17}
Beyer, A. \textit{et al.} The Rydberg constant and proton size from atomic hydrogen. \textit{Science} \textbf{358}, 79-85 (2017)

\bibitem{Fleurbaey18}
Fleurbaey, H. \textit{et al.} New measurement of the $1S - 3S$ transition frequency of hydrogen: contribution to the proton charge radius puzzle. \textit{arXiv}:1801.08816 (2018)

\bibitem{Pohl16}
Pohl, R. \textit{et al.} Laser spectroscopy of muonic deuterium. \textit{Science} \textbf{353}, 669-673 (2016)

\bibitem{Nebel12}
Nebel, T. \textit{et al.} The Lamb-shift experiment in Muonic helium. \textit{Hyperfine Interact} \textbf{212}, 195-201 (2012)

\bibitem{Kandula10}
Kandula, D. Z., Gohle, C., Pinkert, T. J., Ubachs, W. M. G. and Eikema, K. S. E. Extreme Ultraviolet Frequency Comb Metrology. \textit{Phys. Rev. Lett.} \textbf{105}, 063001 (2010)

\bibitem{Hodgman09}
Hodgman, S. S. \textit{et al.} Metastable Helium: A New Determination of the Longest Atomic Excited-State Lifetime. \textit{Phys. Rev. Lett.} \textbf{103}, 053002 (2009)

\bibitem{CancioPastor04}
Cancio Pastor, P. \textit{et al.} Absolute Frequency Measurements of the $2 \ ^3S_1 \rightarrow 2 \ ^3P_{0,1,2}$ Atomic Helium Transitions around 1083 nm. \textit{Phys. Rev. Lett.} \textbf{92}, 023001 (2004)

\bibitem{Luo13}
Luo, P.-L., Peng, J-L, Shy, J.-L. and Wang, L.-B. Precision Frequency Metrology of Helium $2 \ ^1S_0 \rightarrow 2 \ ^1P_1$ Transition. \textit{Phys. Rev. Lett.} \textbf{111}, 013002 (2013)

\bibitem{Noter14a}
Notermans, R. P. M. J. W. and Vassen, W. High-Precision Spectroscopy of the Forbidden $2 \ ^3S_1 \rightarrow 2 \ ^1P_1$ Transition in Quantum Degenerate Metastable Helium. \textit{Phys. Rev. Lett.} \textbf{112}, 253002 (2014)

\bibitem{Luo16}
Luo, P.-L. \textit{et al.} Precision frequency measurements of $\mathrm{^{3,4}He}$ $2 \  ^3P \rightarrow 3 \ ^3D$ transitions at 588~nm. \textit{Phys. Rev. A} \textbf{94}, 062507 (2016)

\bibitem{Zheng17b}
Zheng, X. \textit{et al.} Measurement of the Frequency of the $2 \ ^3S - 2 \  ^3P$ Transition of $^4$He. \textit{Phys. Rev. Lett.} \textbf{119}, 263002 (2017)

\bibitem{Huang18}
Huang, Y.-J. \textit{et al.} Frequency measurement of the $2 \ ^1 S_0 - 3 \ ^1 D_2$ two-photon transition in atomic $^4$He. \textit{Phys. Rev. A} \textbf{97}, 032516 (2018)

\bibitem{Borbely09}
Borbely, J. S. \textit{et al.} Separated oscillatory-field microwave measurement of the $2 \ ^3P_1 - 2 \ ^3P_2$ fine-structure interval of atomic helium. \textit{Phys. Rev. A} \textbf{79}, 060503(R) (2009)

\bibitem{Smiciklas10}
Smiciklas, M. and Shiner, D. Determination of the Fine Structure Constant Using Helium Fine Structure. \textit{Phys. Rev. Lett.} \textbf{105}, 123001 (2010)

\bibitem{Zheng17a}
Zheng, X. \textit{et al.} Laser Spectroscopy of the Fine-Structure Splitting in the $2 \ ^3P_J$ Levels of $^4$He. \textit{Phys. Rev. Lett.} \textbf{118}, 063001 (2017)

\bibitem{Marsman15}
Marsman, A. Horbatsch, M. and Hessels, E. A. Quantum interference effects in saturated absorption spectroscopy of n = 2 triplet helium fine structure. \textit{Phys. Rev. A} \textbf{91}, 062506 (2015)

\bibitem{Wang04}
Wang, L.-B. \textit{et al.}
Laser Spectroscopic Determination of the $^6$He Nuclear Charge Radius. \textit{Phys. Rev. Lett.} \textbf{93}, 142501 (2004)

\bibitem{Mueller07}
Mueller, P. \textit{et al.}
Nuclear Charge Radius of $^8$He. \textit{Phys. Rev. Lett.} \textbf{99}, 252501 (2007)

\bibitem{Shiner95}
Shiner, D., Dixson, R. and Vedantham, V. Three-Nucleon Charge Radius: A Precise Laser Determination Using $^3$He. \textit{Phys. Rev. Lett.} \textbf{74}, 183553 (1995)

\bibitem{CancioPastor12}
Cancio Pastor, P. \textit{et al.} Frequency Metrology of Helium around 1083 nm and Determination of the
Nuclear Charge Radius. \textit{Phys. Rev. Lett.} \textbf{108}, 143001 (2012)

\bibitem{vRooij11}
van Rooij, R. \textit{et al.} Frequency metrology in quantum degenerate helium: Direct measurement of the $2 \ ^3S_1 \rightarrow 2 \ ^1S_0$ transition. \textit{Science} \textbf{333}, 196 (2011)

\bibitem{Sick14}
Sick, I. Zemach moments of $\mathrm{^3 He}$ and $\mathrm{^4 He}$. \textit{Phys. Rev. C} \textbf{90}, 064002 (2014)

\bibitem{Noter14b}
Notermans, R. P. M. J. W., Rengelink, R. J., van Leeuwen, K. A. H. and Vassen, W. Magic wavelengths for the $2 \ ^3S \rightarrow 2 \ ^1S$ transition in helium. \textit{Phys. Rev. A} \textbf{90}, 052508 (2014)

\bibitem{Campbell17}
Campbell, S. L. \textit{et al.} A Fermi-degenerate three-dimensional optical lattice clock. \textit{Science} \textbf{358}, 90-94 (2017)

\bibitem{Marti18}
Marti, G. E. \textit{et al.} Imaging optical frequencies with 100~$\mu$Hz precision and 1.1~$\mu$m resolution. \textit{Phys. Rev. Let.} \textbf{120}, 103201 (2018)

\bibitem{Ludlow15}
Ludlow, A., Boyd, M. M., Ye, J., Peik, E. and Schmidt P. Optical atomic clocks. \textit{Rev. Mod. Phys.} \textbf{87}, 637-701 (2015)

\bibitem{Mitroy13}
Mitroy, J. and Tang, L.-Y. Tune-out wavelengths for metastable helium. \textit{Phys. Rev. A} \textbf{88}, 052515 (2013)

\bibitem{Zhang16}
Zhang, Y.-H., Tang, L.-Y., Zhang, X.-Z. and Shi, T.-Y. Tune-out wavelength around 413 nm for the helium 2 $^3 S_1$ state including relativistic and finite-nuclear-mass corrections. \textit{Phys. Rev. A} \textbf{93}, 052516 (2016)

\bibitem{Henson15}
Henson, B. M. \textit{et al.} Precision Measurement for Metastable Helium Atoms of the 413 nm Tune-Out
Wavelength at Which the Atomic Polarizability Vanishes. \textit{Phys. Rev. Lett.} \textbf{115}, 043004 (2015)

\bibitem{Wu18}
Wu, F.-F. \textit{et al.} Relativistic full-configuration-interaction calculations of magic wavelengths for the $2 \ ^3 S_1 \rightarrow 2 \ ^1 S_0$ transition of helium isotopes. \textit{arXiv}:1804.01218 (2018)

\bibitem{Fried98}
Fried, D. G. \textit{et al.} Bose-Einstein Condensation of Atomic Hydrogen. \textit{Phys. Rev. Lett.} \textbf{81}, 3807 (1998)

\bibitem{Killian98}
Killian, T. C. \textit{et al.} Cold Collision Frequency Shift of the 1S\textendash 2S Transition in Hydrogen. \textit{Phys. Rev. Lett.} \textbf{81}, 3811 (1998)

\bibitem{Killian00}
Killian, T. C. 1S-2S spectrum of a hydrogen Bose-Einstein condensate. \textit{Phys. Rev. A} \textbf{61}, 033611 (2000)

\bibitem{Noter16}
Notermans, R. P. M. J. W., Rengelink, R. J. and Vassen, W. Comparison of spectral linewidths for quantum degenerate bosons and fermions. \textit{Phys. Rev. Lett.} \textbf{117}, 213001 (2016)

\bibitem{Vassen12}
Vassen, W. \textit{et al.} Cold and trapped metastable noble gases. \textit{Rev. Mod. Phys.} \textbf{84}, 175-210 (2012)

\bibitem{Rengelink16}
Rengelink, R. J., Notermans, R. P. M. J. W. and Vassen, W. A simple 2~W continuous-wave laser system for trapping ultracold metastable helium atoms at the 319.8~nm magic wavelength. \textit{Appl. Phys. B} \textbf{122}, 122 (2016)


\bibitem{Moal06}
Moal, S. \textit{et al.} Accurate Determination of the Scattering Length of Metastable Helium Atoms Using Dark Resonances between Atoms and Exotic Molecules. \textit{Phys. Rev. Lett.} \textbf{96}, 023203 (2006)



\bibitem{Morton06}
Morton, D. C., Wu, Q. and Drake, G. W. F. Energy levels for the stable isotopes of atomic helium ($^4$He I and $^3$He I). \textit{Can. J. Phys.} \textbf{83}, 83-105 (2006)

\bibitem{Mueller91}
M\"uller, M. W. \textit{et al.} Experimental and theoretical studies of the Bi-excited collision systems $\mathrm{He^*}(2 \ ^3 S) + \mathrm{He^*}(2 \ ^3 S, 2 \ ^1 S)$ at thermal and subthermal kinetic energies. \textit{Z. Phys. D} \textbf{21} 89-112 (1991)

\bibitem{Diepold16}
Diepold, M. \textit{et al.} Theory of the lamb shift and fine structure in muonic $^4$He ions and the muonic $^3$He-$^4$He isotope shift. \textit{arXiv}:1606.05231v2 (2017)

\bibitem{Borbely12}
Borbely, J. S., van Rooij, R., Knoop, S. and Vassen, W. Magnetic-field-dependent trap loss of ultracold metastable helium. \textit{Phys. Rev. A}, \textbf{85}, 022706 (2012)

\bibitem{Grimm00}
Grimm, R., Weidem\"uller, M., and Ovchinnikov, Y. B. Optical Dipole Traps for Neutral Atoms. \textit{Adv. in atomic, molecular and optical Physics} \textbf{42}, 95-170 (2000)

\bibitem{BIPMcircT}
Bureau international des poids et mesures (BIPM) Circular T
bulletin. \textbf{URL} \url{ https://www.bipm.org/en/bipm-services/
timescales/time-ftp/Circular-T.html } (2018)

\end{thebibliography}

\end{document}


\maketitle
\section{vector and tensor polarizability}
The measurement of the magic wavelength is more complicated when the tensor and vector polarizability are considered. The measured magic wavelength then depends on the details of the magnetic substates that are measured, on the polarization of the laser light and the orientation of the quantization axis. However, the tensor and vector polarizabilities are sufficiently small that these details do not affect the measured values at the current level of uncertainty.

First of all, the vector and tensor polarizabilities do not affect the frequency metrology in any way. The tensor polarizability drops out as common mode because we exclusively use the $m_J = \pm1$ magnetic substates which are affected identically. The vector polarizability introduces an asymmetry between the $m_J=+1$ and $m_J=-1$ substates in the presence of an excess of either $\sigma^+$ or $\sigma^-$ polarization but this asymmetry extrapolates to zero in the trap ac-Stark shift regression that is performed. The non-scalar terms in the polarizability are therefore only relevant to the measurement of the magic wavelength.

In general, the polarizability $\alpha$ in a laser field is given by~\cite{Derevianko11}
\begin{equation}
\alpha(\omega) = \alpha^S_J(\omega) + (\hat{k} \cdot \hat{B}) q \frac{m_J}{2J} \alpha^V_J(\omega) + (3 | \hat{\epsilon} \cdot \hat{B}|^2 -1) \frac{3 m_J^2 -J(J+1)}{2J(2J-1)} \alpha^T_J(\omega),
\label{eq:polarizability}
\end{equation}
where $\alpha^S_J$, $\alpha^V_J$, and $\alpha^T_J$ are, respectively, the scalar, vector, or tensor part of the polarizability of a state with angular momentum $J$, the unit vectors $\hat{k}$, $\hat{B}$, and $\hat{\epsilon}$ are in the direction of, respectively, light propagation, the magnetic field, and the electric field polarization of the light. $q$ indicates the degree of circular polarization, with 0 for linear polarization, and $\pm 1$ for $\sigma^\pm$.

In order to estimate the vector and tensor polarizabilities, we have extended the calculations of the polarizability in ref.~\cite{Noter14} to include the $| 2 \ ^3 S_1 ; m_J=0 \rangle$ state for $\pi$-polarized light and of the $| 2 \ ^3 S_1 ; m_J=+1 \rangle$ state for both $\sigma^+$ and $\sigma^-$ polarization. From these we calculate the vector and tensor polarizability according to equation~\ref{eq:polarizability}, as
\begin{equation}
\alpha^V = \alpha(m_J=+1,\sigma^-)-\alpha(m_J=+1,\sigma^+),
\end{equation}
and 
\begin{equation}
\alpha^T = \frac{\alpha(m_J=+1,\pi)-\alpha(m_J=0,\pi)}{3}.
\end{equation}
At the magic wavelength we find 0.09 $a_0^3$ and 0.03 $a_0^3$ for the vector and tensor polarizabilities respectively. These numbers are small compared to the scalar polarizability of 189.3 $a_0^3$, and the 0.5 $a_0^3$ uncertainty on the dc-offset on the polarizability curve as found in the magic wavelength measurement.

Moreover, the effects of both the vector and tensor polarizability tend to average out in the geometry of our trap. Nominally, the trap laser polarization is linear in both beams with the polarization of the returning beam orthogonal to the first. This will create a rapidly oscillating polarization gradient over the trap, so that any polarization dependent terms average out. 

\section{Mean-field lineshape}
The lineshape due to the mean-field shift is given by equation~25 of ref.~\cite{Killian00}, and can be written as
\begin{equation}
S(\Delta \nu) = A N \frac{\Delta \nu}{\delta \nu^2_{max}} \sqrt{1-\frac{\Delta \nu}{\delta \nu_{max}}},
\label{eq:Kill}
\end{equation}
where $A$ is a pre-factor not important to the discussion here, $N$ is the number of atoms in the BEC, and $\Delta \nu$ is the detuning from resonance. The profile is equal to zero outside the interval $[0,\delta \nu_{max}]$. The profile's full width is given by
\begin{equation}
\delta \nu_{max} = \frac{4 \pi \hbar^2 (a_{ts} - a_{tt})}{h m} n(0) = \frac{a_{ts} - a_{tt}}{h a_{tt}} \mu, 
\end{equation} 
where $a_{tt}$ and $a_{ts}$ are the triplet-triplet and triplet-singlet scattering lengths respectively, $n(0)$ is the peak density of the BEC, and $\mu$ is the chemical potential of the BEC. A number of factors of two are different between equation~\ref{eq:Kill} and equation~25 of ref.~\cite{Killian00} because we use a one-photon, rather than a two-photon transition.

\begin{figure}[h]
\centering
\includegraphics[width= 0.5 \textwidth]{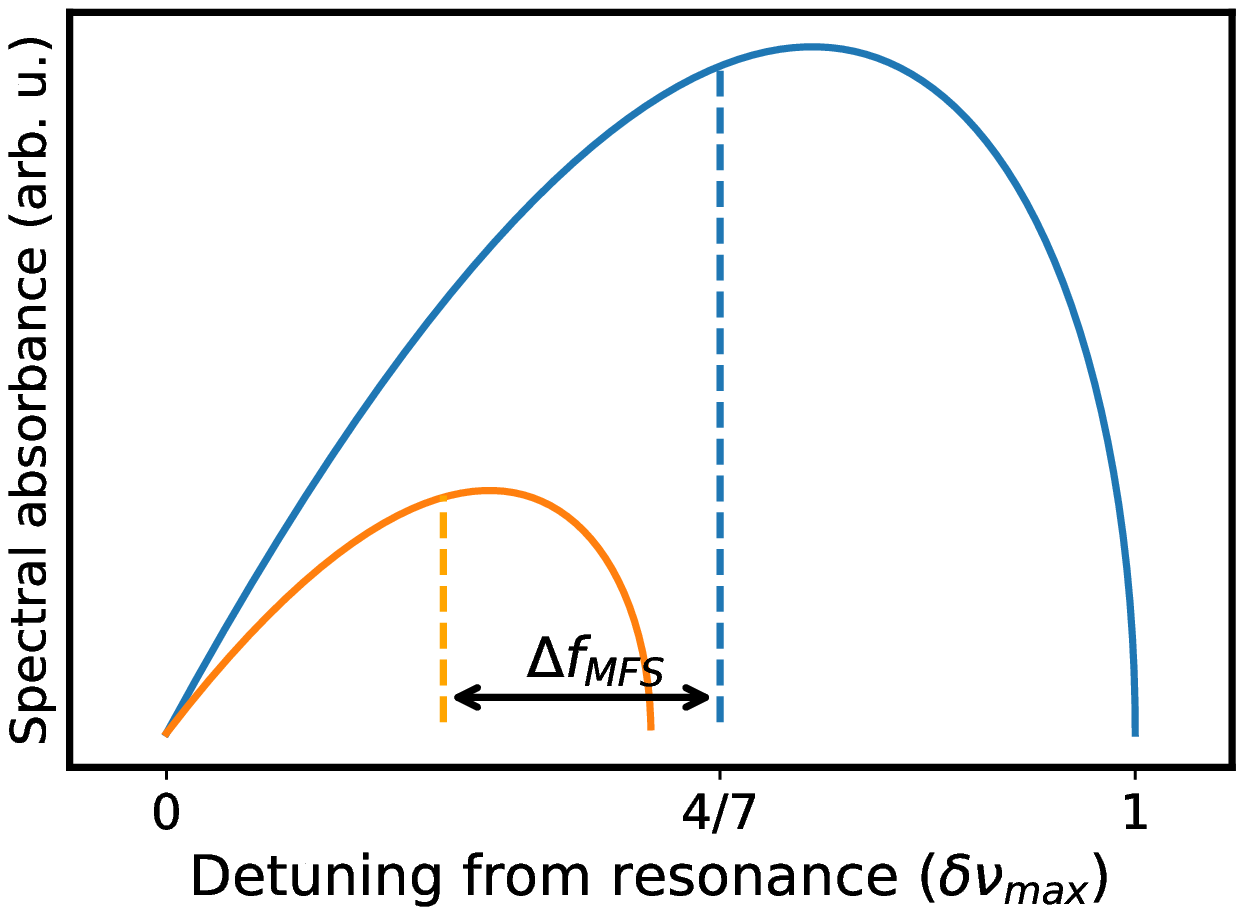}
\caption{Lineshape of a BEC as a result of mean-field interactions according to equation~\ref{eq:Kill}. The average shift of the lineshape is analytically found to be $4/7$ of the full width. The orange line indicates the same line profile with a 50\% smaller chemical potential, giving rise to an observed shift of $\Delta f_{MFS}$.}
\label{fig:MFSdiag}
\end{figure}

The average frequency shift of the lineshape can be calculated according to
\begin{equation}
\langle \Delta \nu \rangle = \frac{\int_0^{\delta \nu_{max}} \Delta \nu S(\Delta \nu) d \Delta \nu}{\int_0^{\delta \nu_{max}} S(\Delta \nu) d \Delta \nu}.
\label{eq:avgmfs}
\end{equation}
The two integrals can be solved using the simple substitution $x = \frac{\Delta \nu}{\delta \nu_{max}}$:
\begin{align}
\int_0^{\delta \nu_{max}} \Delta \nu S(\Delta \nu) d \Delta \nu &= \frac{A N}{\delta \nu_{max}} \int_0^1 x^2 \sqrt{1-x} dx = \frac{16 A N}{105 \delta \nu_{max}} \\
\int_0^{\delta \nu_{max}} S(\Delta \nu) d \Delta \nu &= \frac{A N}{\delta \nu^2_{max}} \int_0^1 x \sqrt{1-x} dx = \frac{4 A N}{ 15 \delta \nu^2_{max}}.
\end{align}
We plug these expressions back into equation~\ref{eq:avgmfs} to find
\begin{equation}
\langle \Delta \nu \rangle = \frac{4}{7} \delta \nu_{max}
\label{eq:avgmfscalc}
\end{equation}
Figure~\ref{fig:MFSdiag} shows the lineshape according to equation~\ref{eq:Kill}, along with the average shift according to equation~\ref{eq:avgmfscalc}.

\section{Dc-Stark shift and blackbody radiation shift}
The dc-Stark shift is given by~\cite{Angel68}:
\begin{equation}
\Delta \nu = -\frac{1}{2 h} \left< \vec{p} \cdot \vec{E} \right> = - \frac{1}{2 h} \mathrm{Re} (\Delta \alpha) E^2,
\end{equation}
where $\vec{p}$ is the instantenous dipole moment, and $\vec{E}$ the electric field. We know the difference in dc-polarizability between the $2 \ ^3 S_1$ and $2 \ ^1 S_0$ states quite well: $\mathrm{Re} (\Delta \alpha) = \left[483.55 \  a_0^3 \right]_{a.u.} \approx 7.97 \times 10^{-39} \ \mathrm{J \ V^{-2} \ m^2}$~\cite{Noter14}. This gives a dc-Stark shift of
\begin{equation}
\frac{\Delta \nu}{E^2} = \frac{4 \pi \epsilon_0 \times 483.55 \ a_0^3}{2 h} \approx 6.02 \times 10^{-6} \  \mathrm{Hz \ V^{-2} \ m^2}.
\end{equation}

Based on a simulation of the electric field in the vacuum chamber using Comsol Multiphysics, we estimate that the ion-MCP produces a field of about 30~V/m at the position of the atoms. Even if we scale this up by a factor ten to be conservative the shift is smaller than 1~Hz.

This analysis also applies for the blackbody radiation (BBR) shift. The mean squared electric field from a blackbody is $\langle E^2 \rangle_T = (831.9 \ \mathrm{ V  \ m^{-1}})^2 \times \left( \frac{T}{300 \ \mathrm{K}} \right)^4$\cite{Ludlow15}, which gives a BBR shift of 4.17~Hz at 300~K. This estimate does not include the so-called dynamic contribution but this is typically smaller than 10\%~\cite{Bloom14}.
\section{Quantum interference}
Quantum mechanically, when multiple paths are available to reach the same final state, the question of which path was taken is ill-defined. In spectroscopy this can lead to frequency shifts due to off-resonant excitation of distant transitions. This is commonly referred to as ``quantum interference" and was realized to be a significant problem for precision spectroscopy on the $2 \ ^3 S \rightarrow 2 \ ^3 P$ transitions~\cite{Horbatsch10}. 

In our work, we observe a transition between the $2 \ ^3 S$ and $2 \ ^1 S$ states. The latter has a natural lifetime of 20~ms in which it decays to the $1 \ ^1 S$ ground state by the emission of two photons. Because we measure (via Penning ionization) the generated population of $2 \ ^1 S$ atoms, quantum interference may occur with transitions to levels that decay to this state. By far the most important of these is the $2 \ ^1 P$ state which decays into the $2 \ ^1 S$ state with a branching ratio of about 0.1\%. To estimate a possible quantum interference shift, we apply the analysis of ref.~\cite{Horbatsch11}, which directly maps to our system when $|1 \ ^1 S\rangle, |2 \ ^3 S\rangle, |2 \ ^1 S \rangle, |2 \ ^1 P \rangle \rightarrow |0\rangle,|1\rangle,|2\rangle,|3\rangle$.

In the limit of short interaction time compared to the Rabi frequency of the main transition ($\Omega_2 T \ll 1$), we find that according to equation 21 of ref.~\cite{Horbatsch11}, the transition frequency is shifted by
\begin{equation}
S_0 = \frac{\gamma_2 \gamma_3 + | \Omega_3 |^2}{4 \omega_{32}},
\label{eq:qi}
\end{equation}
where $\gamma_i$ is the inverse lifetime, and $\Omega_i$ the Rabi frequency for level $|i\rangle$, and $\omega_{ij}$ is the angular frequency difference between levels $|i\rangle$ and $|j\rangle$. It is only the first term in the numerator that corresponds to quantum interference, the second term is simply the ac-Stark shift from this particular level, which is already corrected for by the ac-Stark shift extrapolation. 

$\gamma_3$ and $\omega_{32}$ are known to be $2 \pi \times 287 \ \mathrm{MHz}$ and $2 \pi \times 146 \ \mathrm{THz}$ respectively~\cite{Notermans14}. For $\gamma_2$ there is some ambiguity: the natural linewidth is $2 \pi \times 8 \ \mathrm{Hz}$ but the lifetime of this state is severely reduced because of Penning ionizing collisions with the remaining $2 \ ^3 S$ atoms. The inverse of this lifetime however, cannot exceed the observed linewidth, indicating that $\gamma_2 < 2 \pi \times 10 \ \mathrm{kHz}$. In the interest of providing a conservative estimate, we will use this latter value since it gives the largest shift.

Evaluating equation~\ref{eq:qi} with these numbers gives a shift of approximately $2 \pi \times 4.9 \ \mathrm{mHz}$. However, the asumption $\Omega_2 T \ll 1$ is not justified, and a more thorough calculation is needed. We calculate the Rabi frequency according to
\begin{equation}
| \Omega_2 |^2 = \frac{6 \pi c^2}{\hbar \omega_{12}^3} A_{12} | \langle j_1 m_1 j_2 m_2 |J M \rangle |^2 I_0,
\end{equation}
where $\omega_{12} \approx 2 \pi \times 192.5 \ \mathrm{THz}$ is the 
transition frequency, $I_0 \approx 5.5 \times 10^5 \ \mathrm{ W \ m^{-2}} $ is the peak intensity of the laser light, $A_{12} = 9.1 \times 10^{-8} \ \mathrm{s^{-1}}$ is the Einstein A-coefficient, and $\langle j_1 m_1 j_2 m_2 |J M \rangle \leq 1/\sqrt{3}$, depending on the polarization of the laser light, is the Clebsch-Gordan coefficient of the transition. From these numbers we calculate a Rabi-frequency that we round up to $2 \pi \times 100 \mathrm{Hz}$. Using this number we calculate the expected line profile from equation~25 of ref.~\cite{Horbatsch11}, which has a slightly larger quantum interference shift of $2 \pi \times 79.6 \ \mathrm{mHz}$. This number should be understood as an absolute upper bound, the true shift is likely to be orders of magnitude smaller. For reference, table~\ref{tab:parms} shows all the parameters used to arrive at this estimate.

\begin{table}[h]
\centering
\caption{Parameters used to calculate the upper bound estimate shift due to quantum interference. We take $\Omega_3$ as a multiple of $\Omega_2$ according to $\Omega_3 / \Omega_2 = \sqrt{A_{13}/A_{12}}$ since they are derived from the same laser beam. Using these parameters, we find a shift of $\sim 2 \pi \times 80 \ \mathrm{mHz}$.}
\begin{tabular}{|l|r|r|}
\hline
\textbf{Parameter} & \textbf{value} & \textbf{Remarks} \\ 
\hline
$\Omega_2$ & $2 \pi \times 100 \ \mathrm{Hz}$ & Upper bound \\
$\Omega_3/\Omega_2$ & $4 \times 10^3$ & Same laser beam \\
$\gamma_2$ & $2 \pi \times 10 \ \mathrm{kHz}$ & Upper bound \\
$\gamma_{2 \rightarrow 0}$ & $2 \pi \times 8 \ \mathrm{Hz}$ & \\
$\gamma_3$ & $2 \pi \times 287 \ \mathrm{MHz}$ & \\
$\omega_{23}$ & $2 \pi \times 146 \ \mathrm{THz}$ & \\
T & 100~ms & \\
\hline
\end{tabular}
\label{tab:parms}
\end{table}
\section{Second-order Zeeman shift}
The second-order Zeeman shift of the $2 \ ^1 S_0$ and $2 \ ^3 S_1$ levels are given in~\cite{Vassen16} at 3.2~$\mathrm{mHz \ G^{-2}}$ and 2.3~$\mathrm{mHz \ G^{-2}}$ respectively. The ambient magnetic fields present during spectroscopy (typically about 0.5~G) would thus produce a shift $<0.3 \ \mathrm{mHz}$, which is completely negligible.